\documentclass[journal]{IEEEtran}
\ifCLASSINFOpdf
\else
\fi

\usepackage{cite}
\usepackage{amsmath,amssymb,amsfonts}
\usepackage{graphicx}
\usepackage{textcomp}

\usepackage{booktabs}
\usepackage{url}

\usepackage{algorithm}
\usepackage{algorithmic}

\begin{document}
%
%
%
%

\title{Combining Graph Neural Network and Mamba to Capture Local and Global Tissue Spatial Relationships in Whole Slide Images}
\author{Ruiwen Ding, Kha-Dinh Luong, Erika Rodriguez, Ana Cristina Araujo Lemos da Silva, and William Hsu
\thanks{Ruiwen Ding is with the Medical and Imaging Informatics, Department of Radiological Sciences, Department of Bioengineering at, University of California, Los Angeles, CA, 90024, USA (e-mail: dingrw@ucla.edu).}
\thanks{Kha-Dinh Luong is with the Department of Computer Science, University of California, Santa Barbara, CA, 93106, USA (e-mail: vluong@ucsb.edu)}
\thanks{Erika Rodriguez is with the Department of Pathology and Laboratory Sciences, University of California, Los Angeles, CA, 90024, USA (e-mail: erikarodriguez@mednet.ucla.edu)}
\thanks{Ana Cristina Araujo Lemos da Silva is with Federal University of Uberlandia, MG, Brazil (e-mail: anacals@gmail.com)}
\thanks{William Hsu is with the Medical and Imaging Informatics, Department of Radiological Sciences, Department of Bioengineering at, University of California, Los Angeles, CA 90024 USA (e-mail: whsu@mednet.ucla.edu).}}

\maketitle

\begin{abstract}
In computational pathology, extracting spatial features from gigapixel whole slide images (WSIs) is a fundamental task, but due to their large size, WSIs are typically segmented into smaller tiles. A critical aspect of this analysis is aggregating information from these tiles to make predictions at the WSI level. We introduce a model that combines a message-passing graph neural network (GNN) with a state space model (Mamba) to capture both local and global spatial relationships among the tiles in WSIs. The model's effectiveness was demonstrated in predicting progression-free survival among patients with early-stage lung adenocarcinomas (LUAD). We compared the model with other state-of-the-art methods for tile-level information aggregation in WSIs, including tile-level information summary statistics-based aggregation, multiple instance learning (MIL)-based aggregation, GNN-based aggregation, and GNN-transformer-based aggregation. Additional experiments showed the impact of different types of node features and different tile sampling strategies on the model performance. This work can be easily extended to any WSI-based analysis. Code: \url{https://github.com/rina-ding/gat-mamba}.
\end{abstract}

\begin{IEEEkeywords}
Graph neural network, graph attention network, state space model, Mamba, lung cancer, lung adenocarcinoma, progression-free survival, digital pathology
\end{IEEEkeywords}

%
\IEEEpeerreviewmaketitle

\section{Introduction}
\label{sec:introduction}
\IEEEPARstart{L}{ung} cancer results in more than 1.8 million deaths worldwide each year \cite{sung2021global}. Non-small cell lung cancer (NSCLC), one of two lung cancer types, comprises around 85\% of all lung malignancies in the United States \cite{molina2008non}. Early stage (stage I, II by AJCC 8th edition) NSCLC patients are commonly treated with curative resection, but around 30-55\% of them develop disease recurrence within the first five years of surgery \cite{uramoto2014recurrence}. This high recurrence rate reflects a need to identify early-stage NSCLC patients who may be at high risk of recurrence and give them personalized adjuvant therapies after the surgery. The current clinical standard for NSCLC prognosis and treatment planning is the tumor-node-metastasis (TNM) staging system that assesses tumor size, local invasion, and nodal and distance metastases. However, heterogeneous progression-free survival times are commonly observed among patients with identical TNM staging, suggesting this method is insufficient for risk stratification \cite{woodard2016lung}.

Recently, several studies have shown the benefits of using quantitative histomorphologic features derived from H\&E-stained WSIs to predict recurrence or survival in early-stage NSCLC \cite{wang2020computational}. Due to limitations on GPU memory and the relatively large size of WSIs, WSIs are usually split into non-overlapping equal-sized tiles, and a tile aggregation method is needed to render a slide-level prediction. Most studies aggregate tile-level information using summary statistics such as mean to generate a slide or patient-level prediction \cite{ding2022image}\cite{wang2021prognostic}. Some studies utilize multiple instance learning (MIL)-based methods to aggregate the tiles by taking into account the importance of each tile to the prediction \cite{ilse2018attention}\cite{shao2021transmil}. Other studies model the relationship between the tiles in the WSIs using local message-passing GNN-based approaches where each tile is a node and graph convolution operations are applied to capture the connectivity between nodes \cite{ding2022spatially}\cite{wang2021hierarchical}\cite{chen2021whole}. While promising, message-passing GNN-based approaches are limited by the local node neighborhood information aggregation operation, failing to capture the global long-range dependency between the nodes. Several studies have attempted to leverage the global receptive field of transformers and combine them with the message-passing GNNs to capture both local and global relationships between the tiles \cite{zheng2022graph}\cite{sun2023tgmil}. Although transformers can capture the long-range dependencies among the nodes in the graph, they are bottlenecked by their quadratic computational complexity associated with the self-attention mechanism, especially in applications that require large graphs like the ones in computational pathology.

A recent state space model, Mamba, has been shown to not only maintain the ability to capture long-range dependencies among the tokens in a sequence but also be more computationally efficient than the standard transformers \cite{gu2023mamba}. Mamba has been shown to achieve state-of-the-art results on multiple benchmarks related to long-range molecular graphs that, on average, have hundreds or thousands of nodes in each graph \cite{wang2024graph} \cite{dwivedi2022long}. However, its potential in modeling large graphs from computational pathology has not been as widely explored.

In this work, we introduce an integrated message-passing GNN and state space modeling-based progression-free survival prediction pipeline in early-stage LUAD, the most common subtype of NSCLC. The contributions of work are as follows:

1. We leverage an integrated message-passing GNN and state space model in computational pathology. GAT was used as the GNN, and Mamba was used as the state space model to capture both local and global tissue connectivity in the WSI. 

2. We systematically explored the effect of different node features and tile sampling strategies on model performance.

3. Using patients from two publicly available datasets, we performed extensive experiments by comparing our method with baselines and conducting ablation studies to demonstrate the effectiveness of our proposed method.

\begin{figure*}[]
\centering
\includegraphics[width=1\linewidth]{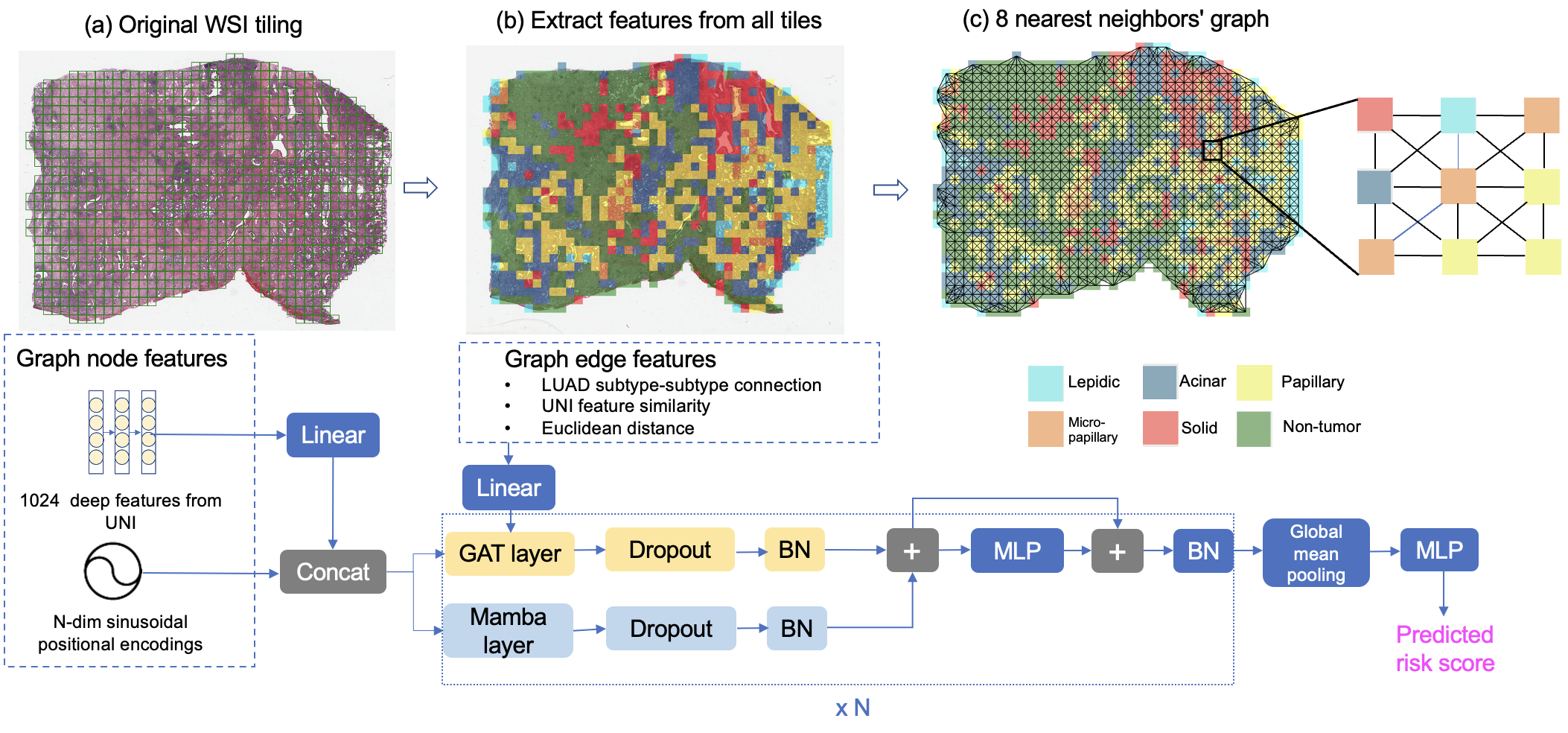}
\caption{The proposed GAT-Mamba pipeline, which consists of WSI tiling (a), node and edge feature extraction (b), graph construction (c), initializing the graph with the extracted node and edge features, and modeling on the graphs. BN: batch normalization. MLP: multi-layer perceptron. +: element-wise summation. N = 16 for positional encodings.} 
\label{pipeline}
\end{figure*}

\section{Related work}
\subsection{Survival analysis in computational pathology}
Most works that utilize artificial intelligence (AI) for cancer prognosis can be divided into two major categories: hand-crafted features-based approaches and deep learning-based approaches. Hand-crafted features are developed using the domain knowledge of pathologists or oncologists \cite{bera2019artificial}. A common approach is to extract quantitative features that describe the shape, texture, and geometric arrangement of all types of nuclei detected on the WSIs \cite{wang2020computational}. Various works have also focused on quantifying the density and spatial arrangement of tumor-infiltrating lymphocytes (TILs) \cite{ding2022image}\cite{abduljabbar2020geospatial}. After extracting the features from each tile, the features are usually aggregated using summary statistics and are fed into a simple linear Cox proportional hazards model to predict the prognosis. While relatively more interpretable and less computationally expensive, these hand-crafted features are usually targeted for a specific cancer or tissue type, limiting their broader utility. 

Compared to hand-crafted features requiring feature engineering, deep learning-based approaches allow one to learn representation from the raw image data directly. Due to the computational complexity in training deep learning models and the lack of fine-grained region-level annotations, many works utilized attention MIL-based approaches. In the context of survival analysis using WSIs, a bag is a WSI, and each tile is an instance. If a WSI is from a high-risk patient, then at least one tile in the WSI must contain malignant tissues; if a WSI is from a low-risk patient, then most or all the tiles must be benign or less malignant. Several works have trained the network to compute the attention score of each tile and aggregate the tile features using weighted pooling \cite{yao2020whole}\cite{ilse2018attention}.

\subsection{Graph-based approaches in computational pathology}
An emerging trend in computational pathology is to model a WSI as a graph and use message-passing GNNs to capture the spatial connectivity between different tissue regions. Message-passing GNNs iteratively aggregate information from the neighboring nodes and update the current node information, and different types of GNNs have different aggregate and update functions. In that way, the GNNs can learn representations that reflect the topological structure of the graph data. For example, Chen et al. \cite{chen2021whole} developed a graph convolutional network (GCN)-based survival analysis pipeline that models each WSI as an 8-nearest-neighbor graph based on the spatial coordinates of the tiles. Ding et al. \cite{ding2022spatially} used a graph isomorphism network (GIN) to predict molecular profiles from WSIs in colon cancer by constructing a graph from the WSIs based on a fixed Euclidean distance threshold derived from the tiles' spatial coordinates. Wang et al. \cite{wang2021hierarchical} constructed a hierarchical graph from both cell-level and tile-level graphs and used a graph attention network (GAT) to predict progression-free survival in prostate cancer. Despite the potentials of GNNs, they have been shown to be prone to the over-smoothing issue where the learned node representations become very similar across nodes after neighborhood information aggregation \cite{wu2024demystifying}. In addition, neighborhood information aggregation is limited to local neighboring nodes, so these models cannot capture the global long-range relationships between the tissue regions.

Few works have attempted to combine GNNs with transformers to alleviate the over-smoothing issue from GNNs and better capture global node connectivity. Zheng et al. \cite{zheng2022graph} devised a pathology image classification pipeline by first passing the WSI-constructed graph into a message-passing GNN and then passing the learned graph representation into a vision transformer. Sun et al. \cite{sun2023tgmil} proposed a hybrid GAT-Transformer model where the WSI-constructed graphs were separately passed into a GAT and a transformer. 

In addition to model architecture, most existing works have not systematically examined different graph construction methods and their influence on the model performance. Most of the graph-based works in computational pathology use either hand-crafted features \cite{wang2021hierarchical} or convolutional neural network (CNN)-extracted features based on either ImageNet transfer learning \cite{ding2022spatially}\cite{chen2021whole}\cite{sun2023tgmil} or self-supervised pretraining \cite{zheng2022graph} on relatively small-scale data. No work has leveraged the recently emerging foundation model features and systematically compared the model's performance when using different node features. Such exploration is important since the performance of a graph model depends on the quality of the input features. Besides the influence of node features on model performance, most existing works have not systematically examined the relationship between tile/node sampling strategy and model performance. Most works build graphs from the WSIs using all available tiles/nodes \cite{zheng2022graph}\cite{chen2021whole}\cite{sun2023tgmil}, and some build graphs by filtering out tiles/nodes from benign or less malignant tissue regions \cite{wang2021hierarchical}\cite{ding2022spatially}. 

\subsection{Long sequence modeling}
Transformers have grown in popularity in both natural language processing and computer vision applications due to their self-attention mechanism that helps model long-range dependency in complex data. However, the transformer's self-attention scales quadratically with respect to the number of tokens in the sequence. Attempts to devise various approximations of the attention mechanism using sparse attention or low-dimensional matrices have been made \cite{zaheer2020big}\cite{choromanski2020rethinking}. However, empirical observations have indicated that these approximations are not ideal for large-scale sequences because these approximations were developed at the expense of the very properties that make transformers effective \cite{gu2023mamba}. Recently, a State Space Model (SSM) named Mamba \cite{gu2023mamba} was proposed to address the computational challenge associated with self-attention in transformers while maintaining superior performance. SSMs can generally be interpreted as a combination of recurrent neural networks (RNNs) and CNNs. To address the inability of SSMs to filter out irrelevant information when updating the sequence embedding, Mamba provides a mechanism for selective inputs (see Model Architecture for details).

\section{Methods}
\subsection{Graph construction}

Figure \ref{pipeline} shows the GAT-Mamba workflow. Each WSI was modeled as a graph. The first step (Figure \ref{pipeline}a) is tiling the original WSI into non-overlapping tiles of either size 512 by 512 at 10x (1 mpp) or 1024 by 1024 at 20x (0.5 mpp) so that the total area covered by a tile is consistent across all patients which have different magnification levels available. Tiles with less than 20\% tissue area were excluded. 

The next steps involve initializing the graph with pre-defined node and edge features (Figure \ref{pipeline}b). There are two groups of node features. The first group is 1024 deep features extracted from a general-purpose self-supervised model for pathology called UNI \cite{chen2024towards}. The second group is an N-dimensional sinusoidal positional encoding feature \cite{dosovitskiy2020image} derived from the relative spatial coordinates of each tile/node within the WSI and N = 16 in this work. In LUAD, patients have one of five predominant histologic subtypes (lepidic, acinar, papillary, micropapillary, solid), each having different prognostic effects on patients. Lepidic is often associated with a better prognosis, acinar and papillary are associated with an intermediate prognosis, and micropapillary and solid are associated with a poorer prognosis \cite{kuhn2018adenocarcinoma}. To capture the heterogeneity of LUAD histologic subtypes and leverage that information in progression-free survival prediction, the subtype-subtype connection between two tiles was used as one of the edge features. There are 21 combinations of subtype-subtype connections among the five subtypes plus non-tumor. In addition, cosine similarity between the UNI-extracted deep features and Euclidean distance between the spatial coordinates of the tiles were used as edge features. 

Similar to \cite{chen2021whole}, each WSI is a graph with tiles being the nodes and the edge connectivity defined by the k-nearest neighbors approach where k = 8 and Euclidean distance between tiles was used as the distance metric (Figure \ref{pipeline}c). The assumption is that the immediate neighboring tiles provide context for each other and potentially share information. 

\subsection{Model architecture}

\begin{algorithm}[t]
\caption{GAT-Mamba}
\label{algo1}
\begin{algorithmic}[1]
  \REQUIRE
  \STATE - $G$ graphs built from $G$ WSIs of a patient. Each $G_i$ has: $N$ nodes and $E$ edges, adjacency matrix $\boldsymbol{A}$ $\in$ $\mathbb{R}^{N \times N}$, $\boldsymbol{X_{UNI}}$ $\in$ $\mathbb{R}^{N \times D_{UNI}}$, $\boldsymbol{X_{PE}}$ $\in$ $\mathbb{R}^{N \times D_{PE}}$, $\boldsymbol{E_{cat}}$ $\in$ $\mathbb{R}^{E \times D_{cat}}$ and $\boldsymbol{E_{cont}}$ $\in$ $\mathbb{R}^{E \times D_{cont}}$ 
    \STATE - $L_{node}$, $EB_{cat}$, $L_{edge}$, $B$ blocks of $GATMambaBlocks$, a global mean pooling layer $Pool$, and a $MLP$ layer.
     \STATE - The event status $event$ and time to event (or follow-up time) $time$ of the patient 
    \STATE - The maximum number of training steps $T$
  \ENSURE Predicted risk score of the patient
\FOR{ $t = 1, ..., T$} 
    \FOR{$g = 1, ..., G$}
        \STATE $\boldsymbol{\hat{X}_{UNI}}$ $\gets$ $L_{node}(\boldsymbol{X_{UNI}})$
        \STATE $\boldsymbol{X}$ $\gets$ Concat($\boldsymbol{\hat{X}_{UNI}}$, $\boldsymbol{X_{PE}}$)
        \STATE $\boldsymbol{\hat{E}_{cat}}$ $\gets$ $EB_{cat} (\boldsymbol{E_{cat}})$, $\boldsymbol{\hat{E}_{cont}}$ $\gets$ $L_{edge} (\boldsymbol{E_{cont}})$
        \STATE $\boldsymbol{E}$ $\gets$ $\boldsymbol{\hat{E}_{cat}} + \boldsymbol{\hat{E}_{cont}}$ 
        \FOR{ $b = 1, ..., B$} 
            \STATE $\boldsymbol{X}$ $\gets$ $GATMambaBlock_b (\boldsymbol{A}, \boldsymbol{X}, \boldsymbol{E})$
        \ENDFOR
        \STATE $\boldsymbol{X}$ $\gets$ $Pool (\boldsymbol{X})$
    \ENDFOR
    \STATE $\boldsymbol{X}$ $\gets$ $MLP (\boldsymbol{X})$
    \STATE Update $L_{node}$, $EB_{cat}$, $L_{edge}$, $GATMambaBlocks$, and $MLP$ by minimizing the Cox loss between the predicted risk score and the observed time to event $time$
\ENDFOR 
\end{algorithmic}
\end{algorithm}

\begin{algorithm}[t]
\caption{Algorithm for $GATMambaBlock$}
\label{algo2}
\begin{algorithmic}[1]
  \REQUIRE
    \STATE - A graph $G$ with $N$ nodes and $E$ edges, adjacency matrix $\boldsymbol{A}$ $\in$ $\mathbb{R}^{N \times N}$, node features $\boldsymbol{X}$ $\in$ $\mathbb{R}^{N \times D_{node}}$, Edge features $\boldsymbol{E}$ $\in$ $\mathbb{R}^{E \times D_{edge}}$
    \STATE - A GAT layer $GAT$, a Mamba layer $Mamba$, a batch normalization layer $BatchNorm$, an MLP layer $MLP$, the hidden dimension of the network, $D$, and number of layers $l$ $\in$ $[1, L]$ 
  \ENSURE The updated node representation $\boldsymbol{X}^L\in\boldsymbol{R}^{N \times D}$
\FOR{ $l = 1, ..., L$} 
\STATE $\boldsymbol{\hat{X}^{l + 1}_{GAT}}$ $\gets$ $GAT^l(\boldsymbol{X}^l, \boldsymbol{E}^l, \boldsymbol{A})$
\STATE $\boldsymbol{X^{l + 1}_{GAT}}$ $\gets$ $BatchNorm (Dropout (\boldsymbol{\hat{X}^{l + 1}_{GAT}}))$ 
\STATE $\boldsymbol{\hat{X}^{l + 1}_{Mamba}}$ $\gets$ $Mamba^l(\boldsymbol{X}^l)$
\STATE $\boldsymbol{X^{l + 1}_{Mamba}}$ $\gets$ $BatchNorm (Dropout (\boldsymbol{\hat{X}^{l + 1}_{Mamba}}))$ 
\STATE $\boldsymbol{X}^{l + 1}_{GATMamba}$ $\gets$ $\boldsymbol{X^{l + 1}_{GAT}} + \boldsymbol{X^{l + 1}_{Mamba}}$
\STATE $\boldsymbol{X}^{l + 1}$ $\gets$ $BatchNorm(MLP^l(\boldsymbol{X}^{l + 1}_{GATMamba}) + \boldsymbol{X}^{l + 1}_{GATMamba})$
\ENDFOR
\end{algorithmic}
\end{algorithm}

\subsubsection{The GAT branch}
In this work, graph convolution operation was performed using GAT \cite{velickovic2017graph}, which uses an attention mechanism to learn the importance of each neighboring node to the current node. This mechanism focuses the network on the most relevant nodes to make predictions. Since edge features are included in this pipeline, they were concatenated with the neighboring ($j$) and current ($i$) node features when computing the attention coefficients $\alpha_{i, j}$:

\begin{equation}\label{gat_eq}
\small
\alpha_{i, j} = \frac{exp(LeakyReLU(\boldsymbol{a}^{T} [\boldsymbol{W}\boldsymbol{X}_{i} \mathbin\Vert \boldsymbol{W}\boldsymbol{X}_{j} \mathbin\Vert \boldsymbol{W}\boldsymbol{E}_{i, j}]))}{\sum_{k \in \mathcal{N}_{i}} exp(LeakyReLU(\boldsymbol{a}^{T} [\boldsymbol{W}\boldsymbol{X}_{i} \mathbin\Vert \boldsymbol{W}\boldsymbol{X}_{k} \mathbin\Vert \boldsymbol{W}\boldsymbol{E}_{i, k}]))}
\end{equation}

where $T$ represents transposition, $\mathbin\Vert$ is concatenation, $\boldsymbol{a}$ is the learnable parameter in the single-layer feedforward neural network that learns the attention coefficients, $\boldsymbol{W}$ is a linear transformation shared across all node and edge features, $\boldsymbol{X}$ is node features, $\boldsymbol{E}$ is edge features, $\mathcal{N}_{i}$ is some neighboring node of current node $i$, and $k$ represents all neighboring nodes $j$ of the current node $i$. Once $\alpha_{i, j}$ is obtained for all neighboring nodes, the current node's features are updated by the weighted sum of its neighboring node features. 

\subsubsection{The Mamba branch}
Similar to RNNs, SSMs map the input sequence $x(t) \in \mathbb{R}^N$ to output sequence $y(t) \in \mathbb{R}^N$ via a hidden state $h(t) \in \mathbb{R}^N$ using a linear ordinary differential equation for continuous input:

\begin{equation}
h^\prime (t) = \boldsymbol{A}h(t) + \boldsymbol{B}x(t), 
\end{equation}
\begin{equation}\label{ssm_eq}
y(t) = \boldsymbol{C}h(t)
\end{equation}

where $\boldsymbol{A}$ is a state matrix that compresses all the past information in the sequence, $\boldsymbol{B}$ is the input matrix, and $\boldsymbol{C}$ is the output matrix. Together, $\boldsymbol{A}h(t)$ represents how the current state evolves over time, $\boldsymbol{B}x(t)$ represents how the input affects the state, and $\boldsymbol{C}h(t)$ represents how the current state translates to output. $\boldsymbol{A}$ and $\boldsymbol{B}$ are discretized using a step size $\Delta$, such that $ \boldsymbol{A} = exp(\Delta \boldsymbol{A})$, $ \boldsymbol{B} = (\Delta \boldsymbol{A})^{-1}(exp(\Delta \boldsymbol{A}) - I)\Delta B$. Both SSMs and its improved Structured State Space Sequence (S4) model \cite{gu2021efficiently} can represent long sequences, but they are limited by their static representation that is not context-aware. That is, matrices $\boldsymbol{A}$, $\boldsymbol{B}$, and $\boldsymbol{C}$ are always constant regardless of the input tokens $x$ in a sequence. To address this issue, Mamba \cite{gu2023mamba} introduced a selection mechanism that allows the model to selectively retain information. Specifically, that is achieved by parameterizing $\boldsymbol{B}$, $\boldsymbol{C}$, and $\Delta$  over the input $x$, where $\boldsymbol{B}$ enables the model to control the influence of input $x_t$ on the hidden state $h_t$ and $\boldsymbol{C}$ enables the model to control the influence of $h_t$ on the output $y_t$ based on the context. $\Delta$ controls how much to focus on or ignore $x_t$, and a larger $\Delta$ means more focus is given to  $x_t$ as compared to the previous hidden states. In this work, each node in the graph is a token, and Mamba's selection mechanism allows the model to minimize the influence of unimportant nodes at each step of hidden state computation. Similar to graph-based transformers, when modeling graph data using Mamba, positional information of the nodes needs to be explicitly encoded into the model since the graph connectivity information is lost when turning the graph into a sequence. In this work, sinusoidal positional encoding \cite{dosovitskiy2020image} serves as the positional information for Mamba. 

\subsubsection{The GAT-Mamba pipeline}
Suppose each patient has up to $G$ WSIs, and each WSI is represented by a graph $G$. Each $G_i$ has $N$ nodes and $E$ edges, and adjacency matrix $\boldsymbol{A}$ $\in$ $\mathbb{R}^{N \times N}$. There are two groups of node features, UNI node features $\boldsymbol{X_{UNI}}$ $\in$ $\mathbb{R}^{N \times D_{UNI}}$, positional encoding node features $\boldsymbol{X_{PE}}$ $\in$ $\mathbb{R}^{N \times D_{PE}}$.  First, a linear layer $L_{node}$ was applied to transform $\boldsymbol{X_{UNI}}$ (Algorithm \ref{algo1} line 7). Then $\boldsymbol{\hat{X}_{UNI}}$ was concatenated with $\boldsymbol{X_{PE}}$ to form the final node features $\boldsymbol{X}$ (Algorithm \ref{algo1} line 8). There are two types of edge features: categorical edge features (subtype-subtype connection) $\boldsymbol{E_{cat}}$ $\in$ $\mathbb{R}^{E \times D_{cat}}$ and continuous edge features $\boldsymbol{E_{cont}}$ $\in$ $\mathbb{R}^{E \times D_{cont}}$. First, an edge embedding function ($EB_{cat}$) was used to transform $\boldsymbol{E_{cat}}$ into continuous features $\boldsymbol{\hat{E}_{cat}}$ and a linear layer ($L_{edge}$) was used to transform $\boldsymbol{E_{cont}}$ into continuous features that share the same dimension as $\boldsymbol{\hat{E}_{cat}}$ (Algorithm \ref{algo1} line 9). The final edge features  $\boldsymbol{E}$ were formed by summing these two groups of transformed edge features element-wise (Algorithm \ref{algo1} line 10). 

The final node features $\boldsymbol{X}$ were passed into the Mamba branch, and both $\boldsymbol{X}$ and $\boldsymbol{E}$ were passed into the GAT branch. The output of the GAT branch was summed with the Mamba branch element-wise (Algorithm \ref{algo2} line 8), and the resulting representation was passed to a global mean pooling layer to generate a patient-level embedding (Algorithm \ref{algo1} line 14) before using an MLP layer to predict the risk score (Algorithm \ref{algo1} line 16). The negative logarithm of Cox partial likelihood loss from DeepSurv \cite{katzman2018deepsurv} was used as the training objective. Cox loss is a ranking loss that penalizes predicted risk scores that are not in concordance with the patients' time to event. Algorithm \ref{algo1} shows each step in the pipeline.

\section{Experiments and Results}
\subsection{Dataset}
Two publicly available datasets, the National Lung Screening Trial (NLST, 132 patients) and the Cancer Genome Atlas (TCGA, 312 patients), were used. For both datasets, all patients had surgical resection as treatment. The inclusion criteria include the patients being stage I or II LUAD, having at least one H\&E-stained WSI, having progression status (either recurrence or lung cancer death), time to event for those who had progression,  follow-up time for those who did not have progression, and for those who had progression, it occurred within five years of surgery. Cases with substantial artifacts, such as pen marks on the WSIs, were excluded. Table \ref{table: patients} summarizes each dataset.

\begin{table}[]
\caption{Patient characteristics}
\centering{
\scalebox{0.9}{
\begin{tabular}{@{}lllll@{}}
\toprule
Traits & NLST & TCGA  & All\\ \midrule
Patients & 132 & 312 & 444 \\
WSIs & 243 & 343 & 586 \\
Average num tiles & 807 & 614 & 691 \\
Progression events & 31 (23.5\%) & 146 (46.8\%) & 177 (39.9\%) \\
Median days to event & 736 (157 - 1637) & 463 (15 - 1778) &  503 (15 - 1778) \\
Median follow-up days & 1387 (54 - 2219) & 609 (11 - 7248) & 839 (11 - 7248)\\
\bottomrule
\end{tabular}
}
}
\label{table: patients}
\end{table}

\subsection{Implementation details}
Five-fold cross-validation, stratified by progression-free survival status, was used. The train-validation-test split for each fold was done at the patient level by combining NLST and TCGA patients during model training (60\%, 266 patients), validation (20\%, 89 patients), and testing (20\%, 89 patients). 

The model had one $GATMambaBlock$, 64 hidden dimensions for the UNI node features $\boldsymbol{X_{UNI}}$, 16 hidden dimensions for the positional encoding features $\boldsymbol{X_{PE}}$, and 16 hidden dimensions for the edge features $\boldsymbol{E}$. The model was trained using batch size 16, learning rate 0.00005, weight decay 0.0001, and dropout 0.3 with Adam optimizer. Early stopping with a tolerance of 5 epochs and a maximum of 200 epochs was used to monitor the validation loss. The model was implemented using Torch Geometric \cite{fey2019fast} and PyTorch 2.0 on NVIDIA-RTX-8000 GPUs. The code of $GATMambaBlock$ was based on the one from \cite{wang2024graph}.

\subsection{Evaluation metrics}
All comparison and ablation experiments use the concordance index (C-index) on the test sets as the primary metric \cite{harrell1996multivariable}. Briefly, C-index calculates the proportion of patients whose predicted risks and progression-free survival times are concordant among all uncensored patients. The secondary metric was the dynamic area under the receiver operating characteristic curve (AUC), which measures the model's ability to distinguish patients who experience events at different time points (1, 3, and 5 years) and those who do not \cite{hung2010estimation}. A paired t-test with multiple comparison correction \cite{benjamini1995controlling} was used as the statistical test to compare model performance. The inference time of each model was calculated by running the model in inference mode using the first fold's test set.

\subsection{Comparison with baseline models}
The effectiveness of our GAT-Mamba model was compared against six baseline models that encompass different state-of-the-art methods to aggregate tile-level features to make a WSI/patient-level prediction. To make a fair comparison, UNI features were used in all models and the models' hyperparameters were tuned on our dataset with the original works' recommended parameters as a reference.

1) Clinical: A baseline clinical Cox model \cite{fox2002cox} that includes age, gender, race, and overall pathological stage as variables. Smoking data was not included since 51\% of patients were missing this information.

2) MLP: UNI features $\boldsymbol{X_{UNI}}$ extracted from each tile and summary statistics, including mean, standard deviation, and range, were used to aggregate tile-level features into patient-level features, which were then fed into a two-layer MLP optimized by Cox loss. 

3) AttentionMIL: A MIL model that aggregates the tile-level UNI features using attention scores learned by the network \cite{ilse2018attention}. 

4) TransMIL: A transformer-based MIL model that uses the Nyström method to approximate the self-attention of the tiles, which were used to aggregate tile-level UNI features to classify WSIs. \cite{shao2021transmil}

5) PatchGCN: A graph-based approach that uses DeepGCN \cite{li2019deepgcns} as the GNN and builds graphs from WSIs using the 8-nearest neighbor approach to predict patient survival. 

6) GPT: A graph transformer model \cite{zheng2022graph} that first passes the WSI-built graph into a GCN \cite{kipf2016semi} followed by a vision transformer to classify WSIs.

\begin{figure}[]
\centerline{\includegraphics[width=\columnwidth]{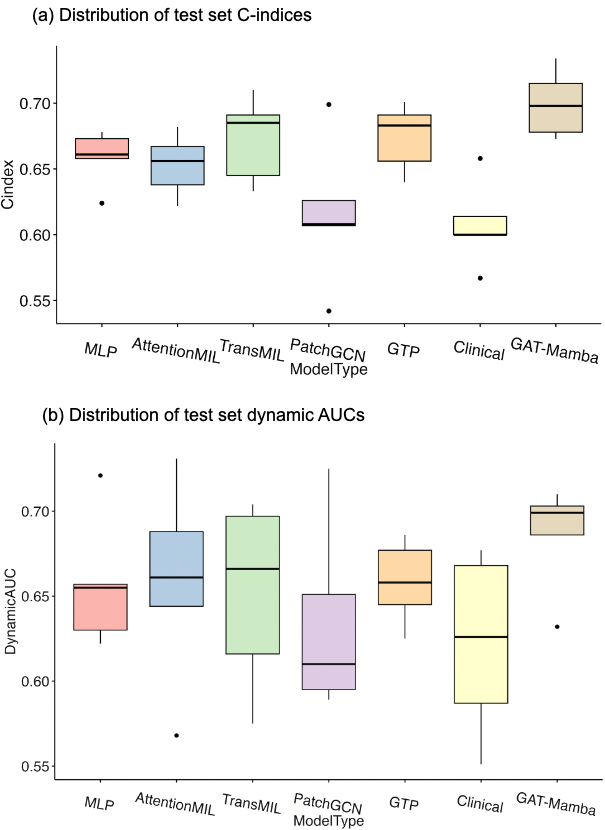}}
\caption{Box plots of the 5-fold cross-validation test set C-indices (a) and dynamic AUCs (b) for GAT-Mamba and all baseline models.}
\label{boxplots}
\end{figure}

As shown in Table \ref{table: baselines} and Figure \ref{boxplots}, GAT-Mamba outperformed all six baseline models regarding both C-index and dynamic AUC. The clinical Cox model was the worst-performing model, resulting in an average C-index of 0.608. PatchGCN was better than the clinical model but worse than the rest. GTP achieved a slightly better average C-index and had a smaller standard deviation than TransMIL. Both GTP and TransMIL had better performance as compared to MLP and AttentionMIL. In addition, as reflected in Figure \ref{boxplots}, GAT-Mamba's performance distribution is relatively less spread out across folds than most baseline models, especially when using dynamic AUCs as the evaluation metric.

\begin{table}[]
\caption{Comparison with baseline models. * p < 0.05 compared to all other models for C-index. InfTime: Inference time in seconds.}
\centering{
\scalebox{0.9}{
\begin{tabular}{@{}llllll@{}}
\toprule
Model   &  Parameters &  InfTime & C-index &  Dynamic AUC  \\ \midrule
MLP    &67,905    &0.0696 & 0.659 $\pm$ 0.0189 & 0.657 $\pm$ 0.0348   \\
AttentionMIL   & 99,202  &111  &  0.653 $\pm$ 0.0211 & 0.658 $\pm$ 0.0539 \\
TransMIL  & 105,041 &100     &  0.673 $\pm$ 0.0291 & 0.652 $\pm$ 0.0493 \\
PatchGCN   & 82,370  &0.422 &  0.616 $\pm$ 0.0502  & 0.634 $\pm$ 0.0504 \\
GTP & 103,633 &0.192 & 0.675 $\pm$ 0.0242 & 0.658 $\pm$ 0.0219 \\
Clinical & 12 &0.003 & 0.608 $\pm$ 0.0331 &  0.622 $\pm$ 0.0478\\
GAT-Mamba* & 127,425 &0.178 & 0.700 $\pm$ 0.0228  & 0.686 $\pm$ 0.0281 \\
\bottomrule
\end{tabular}
}
}
\label{table: baselines}
\end{table}

\subsection{Ablation studies}
The effectiveness of each component of GAT-Mamba was assessed by doing ablation experiments. Specifically, five ablated models were trained and evaluated.

1) GAT: This model only has the GAT branch.

2) Mamba: This model only has the Mamba branch.

3) GAT-Transformer: This model replaces the Mamba branch with a standard transformer.

4) GAT-MambaNo$\boldsymbol{E}$: This model has no edge features as input for the GAT branch.

5) GAT-MambaNo$\boldsymbol{X_{PE}}$: This model does not have positional encoding node features $\boldsymbol{X_{PE}}$.

Table \ref{table: model_alation} shows that all the ablated models performed worse than the complete model. In particular, the complete model was statistically significantly better than the GAT and GAT-Transformer models. 

In addition to modeling ablation, node feature ablation was also performed to assess the impact of different node features on the model performance. 

1) HandCrafted: Studies have found the prognostic values of lymphocytes in the tumor microenvironment of lung cancer \cite{ding2022image}\cite{saltz2018spatial}. In this work, the publicly available HoVer-Net model pretrained on PanNuke dataset was first used for cancer nuclei and lymphocyte detection \cite{graham2019hover}. The detection results were manually verified by a pathologist (E.R.). A total of 57 features related to the density of lymphocytes and the spatial colocalization between lymphocytes and cancer nuclei were extracted from each tile from the WSI\cite{ding2022image}\cite{shaban2019novel}. 

2) LUADDeep: Different histologic subtypes of LUAD have been shown to be associated with patient prognosis \cite{kuhn2018adenocarcinoma}. 512 deep features were extracted from a ResNet18-based pretrained LUAD histologic subtype classifier \cite{ding2023tailoring}. 

3) ResNet50IN: 1024 deep features were extracted from ResNet50 pretrained on ImageNet data \cite{he2016deep}. 

4) CONCH: 512 deep features were extracted from CONCH, a vision language pathology foundation model pretrained on 1.17M image caption pairs \cite{lu2024visual}. 

5) PLIP: 512 deep features were extracted from PLIP, another vision-language pathology foundation model pretrained on over 200k image-text pairs from pathology Twitter \cite{huang2023visual}. 

6) UNI: 1024 deep features were extracted from UNI, a large vision model trained on over 100M pathology images. \cite{chen2024towards}.

Table \ref{table: feature_ablation} shows that GAT-Mamba achieved the best performance when using UNI node features as compared to the other types of features. The pathology foundation model-extracted features (CONCH, PLIP, UNI) and ResNet50IN resulted in better-performing models than the other two feature types, indicating that in general, features extracted from any models that were trained on large datasets (either natural images or pathology images) would be relatively more robust than the ones trained on smaller datasets.

\begin{table}[]
\caption{Modeling ablation study. * means p < 0.05 compared to GAT and GAT-Transformer for C-index, and p < 0.05 compared to GAT for dynamic AUC. InfTime: Inference time in seconds.}
\centering{
\scalebox{0.85}{
\begin{tabular}{@{}lllll@{}}
\toprule
Model   &  Parameters & InfTime & C-index  & Dynamic AUC\\ \midrule
GAT          &  89,666 &0.143 & 0.675 $\pm$ 0.0304 &0.663 $\pm$ 0.0329      \\
Mamba          &  120,401 &0.190 & 0.680 $\pm$ 0.0267 &0.673 $\pm$ 0.0395      \\
GAT-Transformer     &    128,945 &0.735 & 0.676 $\pm$ 0.0430  &  0.678 $\pm$ 0.0636    \\
GAT-MambaNo$\boldsymbol{E}$    & 127,041 & 0.150  &  0.691 $\pm$ 0.0311  & 0.699 $\pm$ 0.0255     \\
GAT-MambaNo$\boldsymbol{X_{PE}}$ & 106,177 &0.211 & 0.687 $\pm$ 0.0447 & 0.680 $\pm$ 0.0684 \\
GAT-Mamba* & 127,425 &0.178 & 0.700 $\pm$ 0.0228 & 0.686 $\pm$ 0.0281 \\
\bottomrule
\end{tabular}
}
}
\label{table: model_alation}
\end{table}

\begin{table}[]
\caption{Node feature ablation study. * means p < 0.05 compared to hand-crafted and LUAD subtype deep features for C-index and dynamic AUC.}
\centering{
\begin{tabular}{@{}lllll@{}}
\toprule
Features  & Raw features & C-index  & Dynamic AUC\\ 
\midrule
HandCrafted      &57    & 0.629 $\pm$ 0.0427  & 0.633 $\pm$ 0.0663     \\
LUADDeep   &512   &0.635 $\pm$ 0.0357  & 0.638 $\pm$ 0.0376     \\
ResNet50IN    &1024   & 0.669 $\pm$ 0.0297  & 0.668 $\pm$ 0.0373     \\
CONCH   &512 &  0.683 $\pm$ 0.0377  &0.663 $\pm$ 0.0463     \\
PLIP &512 & 0.684 $\pm$ 0.0353 &0.671 $\pm$ 0.0396\\
UNI* & 1024 & 0.700 $\pm$ 0.0228 & 0.686 $\pm$ 0.0281 \\
\bottomrule
\end{tabular}
}
\label{table: feature_ablation}
\end{table}

\begin{figure*}[]
\centering
\includegraphics[width=\linewidth]{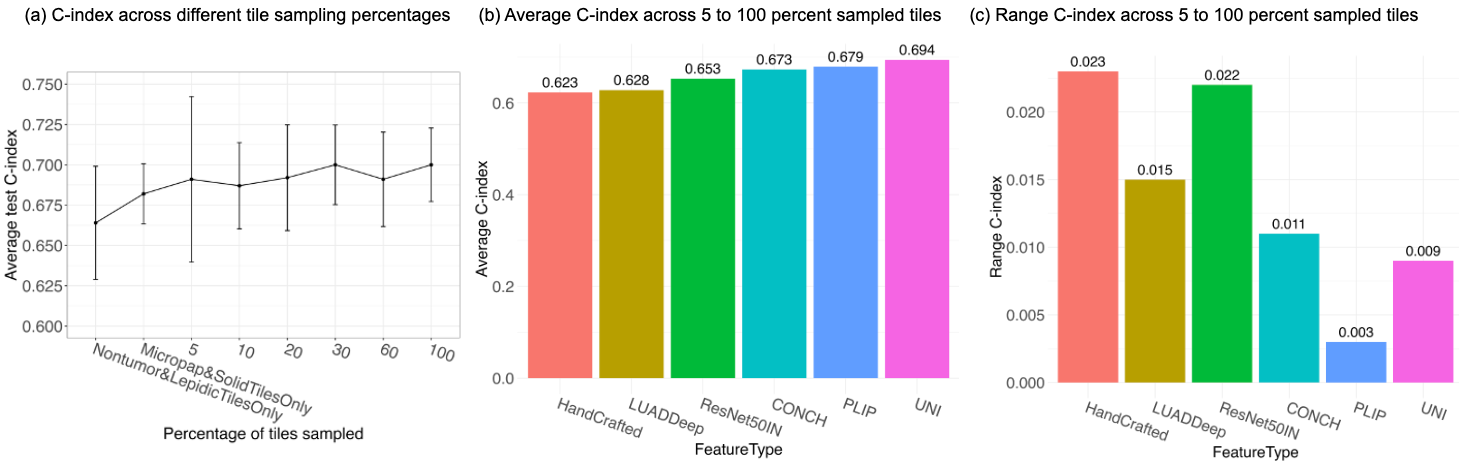}
\caption{Results of tile sampling experiments. (a) represents the line graphs visualizing the average C-index and its standard deviation across different percentages of tiles sampled or when using only aggressive or when using only less aggressive tiles, using UNI node features. (b) represents a bar graph showing the macro-average of the average C-index across 5, 10, 20, 30, 60, and 100 percent sampling for all six types of node features, and (c) represents the macro-range version. }
\label{tile_sampling}
\end{figure*}

\subsection{Different tile sampling strategies}
Since the graphs used in this work were built from the tiles in the WSIs, experiments were done to assess the impact of different tile sampling strategies on the model performance and how the impact would change when using different node features. Specifically, the first kind of tile sampling is based on the aggressiveness of each tile: (1) sampling only the tiles of the most aggressive subtypes (micropapillary and solid) and (2) sampling only the tiles of the least aggressive subtypes (non-tumor and lepidic). The second kind of tile sampling was based on the percentage of tiles (5 to 100 percent), and tiles were randomly selected. These tile sampling strategies were applied to all six groups of node features, respectively.

Figure \ref{tile_sampling} (a) shows the trend of model performance under different sampling strategies when using UNI features as GAT-Mamba node features. Generally speaking, across all node feature types, using 100\% tiles resulted in the best-performing models. However, the amount of performance improvement depends on the type of node features used.  According to Figure \ref{tile_sampling}b, c, generally, when the features are more powerful (better macro-averaged C-index), the model performance will improve less when using more tiles to build the graph. For example, HandCrafted features were the least powerful since they resulted in the lowest average C-index (0.623), and the C-index difference between the best and worst sampling percentages was the largest, meaning this model benefited the most from increasing the number of tiles. However, the model with UNI features did not benefit much from increasing the number of tiles. 

In addition, across all node feature types, sampling only tiles of the least aggressive subtypes or the most aggressive subtypes generally resulted in models that have slightly worse performance than any of the random sampling methods (5 to 100 percent tiles sampling), which implicitly includes tiles of all subtypes. This indicates that including all types of tissue regions in the analysis is important. 

\subsection{Model prediction visualization}
\begin{figure*}[]
\centering
\includegraphics[width=\linewidth]{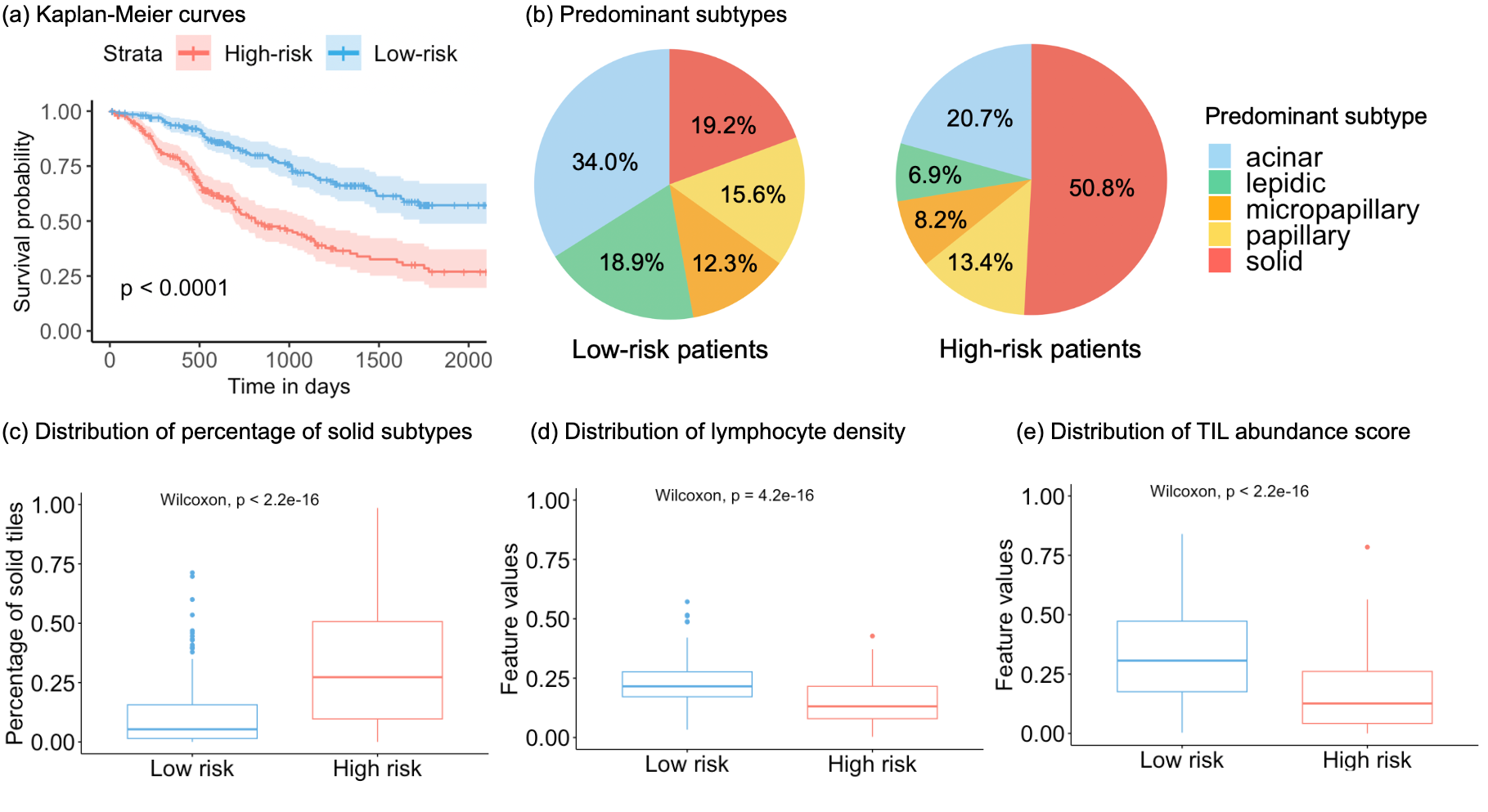}
\caption{Visualization of characteristics of GAT-Mamba predicted risk groups. (a) represents the Kaplan–Meier curves of low and high-risk groups where the log-rank test p-value indicates a statistically significant difference in the progression-free survival distribution of the two groups. (b) represents the overall distribution of predominant histologic subtypes in low (left) and high (right) risk patients. (c) represents the distribution of the percentage of solid tiles in each patient in low and high-risk groups. (d)(e) show the distribution of four hand-crafted features between low and high-risk groups, with (d) representing the number of lymphocytes divided by the number of all nuclei, (e) representing the TIL abundance score \cite{shaban2019novel}. }
\label{visualization}
\end{figure*}

For each of the five models in 5-fold cross-validation, the median risk score from the training set was applied to the test set to get low-risk and high-risk patients. After all test set patients' risk categories were obtained, we examined the characteristics of the two groups. In total, there were 232 high-risk and 212 low-risk patients. Figure \ref{visualization}a shows that the two groups had significantly different progression-free survival distributions. Figure \ref{visualization}b, c indicates that the high-risk patients generally have a more solid subtype as compared to the low-risk patients, and Figure \ref{visualization}b also shows that the low-risk patients tend to have lepidic as the predominant subtype. Figure \ref{visualization}d indicates that low-risk patients have a higher lymphocyte density as compared to high-risk patients, and Figure \ref{visualization}e shows that there is more tumor-lymphocyte colocalization in low-risk patients than high-risk patients. 

\begin{figure*}[]
\centering
\includegraphics[width=\linewidth]{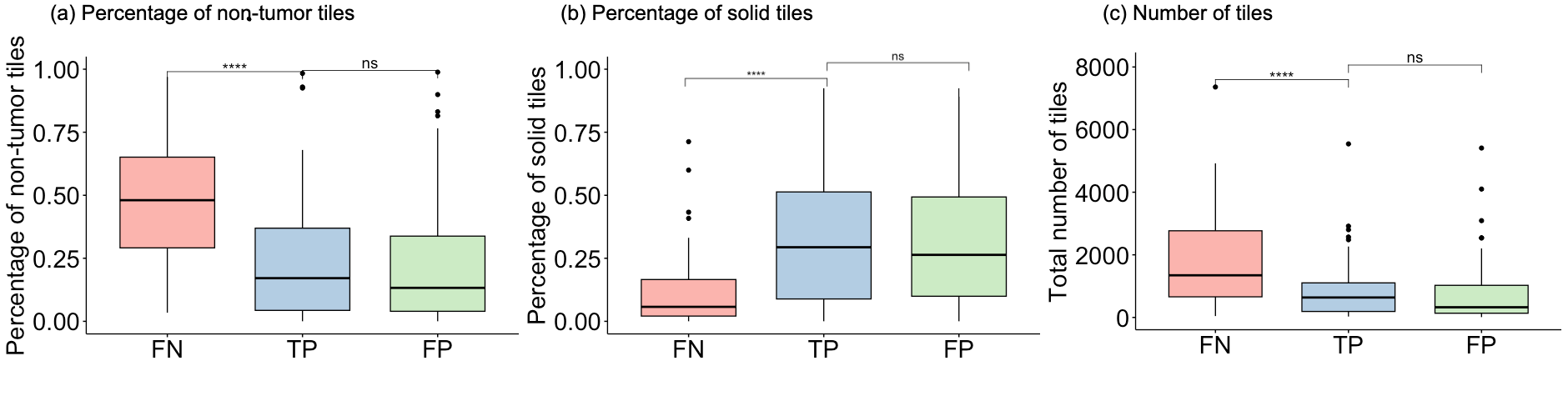}
\caption{Characteristics of tiles for model-predicted false negative (FN), true positive (TP), and false positive (FP) patients. (a) shows the distribution of the percentage of non-tumor tiles. (b) shows the distribution of the percentage of solid tiles. (c) shows the distribution of the total number of tiles. **** means p $<$ 0.001, ns means p $\geq$ 0.05 using Wilcoxon rank sum test.}
\label{error_analysis}
\end{figure*}

\subsection{Error analysis}
The purpose of error analysis in this work is to understand the potential reasons why the model makes a wrong prediction. From a clinical standpoint, false negative predictions are not ideal for prognosis since false negatives mean that a patient potentially would miss opportunities for treatment intervention after underestimating the risk of progression. A true positive is when the patient develops a progression event within 5 years of surgery and has a predicted risk score that falls in the high-risk category. There were a total of 61 (13.7\%) false negatives and 115 (25.9\%) false positives. We found that for false negatives, they had nearly twice as many non-tumor tiles as compared to the true positive patients (Figure \ref{error_analysis}a) and had much less solid tissue than the true positives (Figure \ref{error_analysis}b). We further discovered that, on average, the false negatives had larger tissues as compared to the true positives (Figure \ref{error_analysis}c). Together, these trends indicate that the tissue's malignancy type is more important than the tissue size when it comes to their influence on the model's ability to make a correct prediction for patients with more aggressive disease. There is no obvious trend for false positives.

\section{Discussion}
Due to their large size, WSIs are often divided into small tiles, based on which analysis is performed. A key modeling component is the tile-level information aggregation approach, which gives the aggregated information that can be used to generate a WSI-level prediction. In this work, we proposed a novel pipeline that integrates GAT, a message-passing GNN, with Mamba, a state space model, to capture local and global spatial relationships among the tiles in the WSIs to predict progression-free survival in LUAD patients. 

Our proposed GAT-Mamba outperformed all baseline models and ablations that use different ways to aggregate tile-level features to render a WSI-level prediction and a clinical feature-based model. Experiment results show that the clinical model achieved a C-index of 0.608, but it was inferior to all other models that incorporated WSI-derived features, indicating the potential of adding quantitative image features in improving prognosis. Among the three non-graph-based baseline models, MLP, AttentionMIL, and TransMIL, MLP, and AttentionMIL had similar performance, and TransMIL had the best performance among the three models. This indicates that TransMIL's self-attention mechanism might have helped capture the correlation information between different tissue regions, and such correlation contributed to improved performance, whereas MLP and AttentionMIL did not leverage such correlation since they treated each tile independently. As for message-passing graph-based baselines/ablations, GAT had a much better performance than PatchGCN, indicating that the node aggregation method of GAT might be more effective for this work. The two models that leveraged a combined GNN and transformer achieved very similar results, but they did not outperform GAT alone. In terms of computational efficiency, all models have similar inference times except for attention MIL and TransMIL, which have much longer inference times than the rest. Although GAT-Mamba has more parameters than most other models, its performance and inference time are still advantageous. By considering the trade-offs between model size, inference time, and accuracy, GAT-Mamba is still the most desirable, especially in a high-stakes field like medicine, where accuracy is important.

The reason why GAT-Mamba outperformed the two GNN-transformer-based models is likely related to the input selection mechanism in Mamba. In relatively large graphs like the ones used in this study (691 nodes on average), it is important to filter out noisy nodes that do not contribute valuable information to the prognosis. Instead of compressing all the information like transformers do, Mamba only selectively retains the most informative nodes. The fact that Mamba works better than transformers on pathology graphs also resonates with the findings that Mamba performed better on large long-range molecular data than the transformer-based approaches \cite{wang2024graph}. Combined with the GAT's local neighborhood connectivity modeling function, GAT-Mamba can effectively retain the most informative nodes based on both local and global contexts in the WSI. 

Besides model architecture, different node features were also compared in this study. To date, ResNet50IN-extracted features are one of the most common types of node features in digital pathology graph modeling \cite{ding2022spatially}\cite{chen2021whole}\cite{sun2023tgmil}. Our node feature ablation study shows that pathology foundation model-extracted features PLIP, CONCH, and UNI resulted in better model performance than ImageNet-pretrained-model-extracted features or LUAD domain-specific features. This suggests one can leverage pathology foundation models as feature extractors in their work. UNI features resulted in the best-performing model compared to PLIP and CONCH, which might be because UNI was trained on the most amount of data. In addition, several related works build graphs from the WSIs using all available tiles/nodes \cite{zheng2022graph}\cite{chen2021whole}\cite{sun2023tgmil}. Results from our tile-sampling experiments suggest that one may not need to use all the tiles from the WSI to construct the tile-level graph when robust node features are used. Since WSIs are large and can be broken down into hundreds or thousands of tiles, good node features might help save storage and computational costs associated with bigger graphs.

Although UNI features resulted in the best model performance, they are not as interpretable as other worse-performing features, such as hand-crafted features and histologic subtypes. One way to enhance interpretability is to leverage the more interpretable features in a post-hoc analysis of the model prediction. We found that the model-predicted low-risk patients tend to have a lepidic predominant subtype, whereas the high-risk patients tend to have a solid predominant subtype. This is consistent with the fact that lepidic is the least aggressive subtype, whereas solid is the most aggressive one \cite{kuhn2018adenocarcinoma}. In addition, more immune cells and greater immune-tumor cell colocalization appear in low-risk patients as compared to the high-risk ones, which aligns with our current understanding of lung cancer where immune infiltration is associated with favorable disease outcome \cite{bremnes2016role}\cite{saltz2018spatial}\cite{ding2022image}. Finally, error analysis indicates that the inherent limitation associated with the tissue sampling (a large portion of the benign region) during the surgical resection might have been attributed to some false negative predictions. Although there are still some more malignant regions, they might be insufficient for the model to recognize the aggressiveness of the entire tumor.

Although our proposed GAT-Mamba pipeline showed promising results, we note several areas for improvement. For example, although UNI node features were shown to be the best among all types of node features compared in this study, future work can finetune UNI's model using disease-specific data, which might result in a better node feature extractor than the out-of-box version. Further, the current GAT-Mamba's generalizability to external datasets will likely be limited by the relatively small size of the training data. If larger training data with diverse tissue staining appearances and patient characteristics is available, we can train a larger model with better generalizability. In addition, prognosis is a task that likely requires more than just the pathology image-based information to make a more accurate prediction. Future work will consider combining both pathology and radiology data to get a more comprehensive view of the patient's prognosis.

\section{Conclusion}
In summary, we demonstrated the benefit of combining a message-passing GNN and a state space model Mamba to capture both local and global spatial relationships between different regions in the WSIs. The model outperformed all baselines for predicting progression-free survival in early-stage LUAD patients. Experiments show the impact of different types of node features and different tile sampling strategies on model performance. Although the prediction task of this work is on LUAD prognosis, one can easily use the pipeline on any other digital pathology WSI-based tasks, such as classification. 


%

\ifCLASSOPTIONcaptionsoff
  \newpage
\fi



%
\bibliographystyle{ieeetr}
\bibliography{ref}

%




\end{document}